\documentstyle[11pt]{article}
 
\def\pa{\partial}

 \def\G{\Gamma}
\def\a{\alpha}
\def\b{\beta}
\def\d{\delta} 
\def\e{\epsilon}

\def\l{\lambda} 
\def\m{\mu}
\def\n{\nu}

\def\mn{{\mu\nu}}
\def\be{\begin{equation}}
\def\ee{\end{equation}}
\begin{document}

\begin{center}
{\Large\bf Inequivalence of First and Second Order\\
Formulations in D=2 Gravity Models}\footnote{The present results were reported
in the Proceedings of the Markov Memorial Quantum Gravity Seminar.}

S. Deser\\
Department of Physics\\
Brandeis University, Waltham, MA 02254, USA
\end{center}

\begin{quotation}
The usual equivalence between the Palatini and metric (or affinity and
vielbein) formulations of Einstein theory fails in two spacetime dimensions for
its ``Kaluza--Klein" reduced (as well as for its standard) version.  Among the
differences is the necessary vanishing of the cosmological constant in the
first order forms.  The purely affine Eddington formulation of Einstein theory
also fails here.
\end{quotation}

Most general constructions in physical theories are formally valid in all
dimensions, even though many properties can be quite different in special D.
In this note we present a simple but striking deviation from this rule
in the context of some gravity models at D=2, a dimension that is always ``more
equal" than others.

We shall see that the standard Palatini formulations, in which metric and
affinity or vielbein and connection are independently varied, no longer
coincide with their purely metric or zweibein second order expressions.
While this is known \cite{004} for the usual Einstein Lagrangian, it is also
true for the more interesting, ``Kaluza--Klein reduced", version involving a
Lagrange multiplier \cite{002}.
The source of the difference is the Weyl invariance enjoyed
(only) by the first order Palatini forms. As a result metricity of the initial,
affine, space is no longer recovered from the field equations; the cosmological
constant must also vanish.
We shall also see that even
purely affine expression of Einstein given by Eddington \cite{003} theory is
different in 2D.

We first review the purely Einstein version \cite{004}.
The Einstein--Palatini Lagrangian in any dimension is
\be %1
{\cal L} =  h^{\mn} R_{\mn} (\G ) \; ,\;\;\;\;\;  R_{\mn}(\G ) \; \equiv \;
\G^\a_{\mn ,\a} - \G_{\m ,\n} + \G^\a_{\mn} \G_\a -
\G_{\m\b}^\n \G_{\n\a}^\b
\ee
where $h^{\mn} \equiv \sqrt{-g} \; g^{\mn}$, ~
$\G_\mn^\a   =  \G_{\n\m}^\a , \;\; \G_\m  \equiv \G_{\m\a}^\a$;
here the metric and affinity are to be varied independently.
Commas denote ordinary differentiation and
(since $h^{\mn}$ is symmetric) only the symmetric
part of the Ricci tensor enters in ${\cal L}$.  [In the second
order, purely metric, form (where $\G^a_\mn$ is the Christoffel
symbol) the 2D ${\cal L}$ is a total divergence  -- the Euler
density, but that is not our point here.] In, and only in, 2D, our ${\cal L}$
is manifestly Weyl invariant (if we take $\G^a_\mn$ to be inert)
since the contravariant density $h^\mn$ is unimodular. [One could alternately
take $h^\mn$ to be generic \cite{001}, but such a theory is then not a metric
one at all.]  This new local gauge
freedom will imply that $\G_\m$, whose usual role is to ensure covariant
constancy of $\sqrt{-g}$, is undetermined, {\it i.e.}, that
$\G$-variation of the action will not fix the affinity completely
to be the metric one.  On the other hand, the action's metric variation,
although it does not vanish
identically here, will still turn out to be vacuous just as in second order
form.
The field equations, then, are
\begin{equation}%2
G_\mn \equiv R_\mn (\G ) - \textstyle{\frac{1}{2}} \; g_\mn \;
g^{\a\b}
R_{\a\b} (\G ) = 0 \; , \;\; G^\m_\m \equiv 0
\end{equation}
\begin{equation}%3
D_\a \, h^\mn -  \textstyle{\frac{1}{2}} (\d^\m_\a \, D_\l \,
h^{\l\n}
+ \d^\n_\a \, D_\l \, h^{\l\m} ) = 0
\end{equation}
where the covariant derivatives on the contravariant tensor density $h^\mn$ are
with respect to $\G$, and the
symmetrized part of the $R_\mn$ is  understood in (2).
We also note that Weyl invariance of the Einstein action
forbids a cosmological term $\l\sqrt{-g}$
since the latter does depends on the conformal factor, whose variation
implies that
$\l =0$.  This property is, however, common to second order form
where the trace $G^\m_\m (g)$ vanishes identically as well.

To determine $\G^\a_\mn$, we
first trace (3), which yields $D_\n h^\mn = 0$ and so implies
that
\begin{equation}%4
D_\a h^\mn \equiv \pa_\a h^\mn + \G^\m_{\a\l} h^{\l\n} +
\G^\n_{\a\l} h^{\l\m} -
\G_\a h^\mn  = 0 \; .
\end{equation}
That (4) is not a complete set of equations is clear from the fact
that its $(\mn )$ trace vanishes identically in 2D because $g_\mn \pa_\a h^\mn$
does.
Since $\d h^{\mn} \equiv \sqrt{-g} \, (\d g^\mn - \frac{1}{2} \, g^\mn g_{\a\b}
\d g^{\a\b})$.
Normally, the covariant constancy of the metric or metric density
expressed in (4) does of course determine the affinity completely to be the
metric one. In any D, straightforward algebraic manipulation of (4) yields
\be %5
\G^\a_\mn  =  \{ ^\a_\mn \} \; + \;  \textstyle{\frac{1}{2}}
(g_\mn X^\a - \d^\a_\m X_\n - \d^\a_\n X_\m ) \; , \;\;
X_\m  \equiv  D_\m \;  \ln \; \sqrt{-g} \; = \; \pa_\m \;
\ln \; \sqrt{-g} - \G_\m \;\; .
\ee
whose trace is
\begin{equation}%6
(1-D/2)(\G_\m - \pa_\m \; \ln \; \sqrt{-g} ) = 0 \; .
\end{equation}
Here the dimensionality appears explicitly and hence,
as advertised, spacetime is not entirely fixed to be a purely metric manifold
at D=2, since the $\G_\m$ component of the affinity remains undetermined.

The Einstein equation (2) still remains vacuous.  Inserting (5)
into the Ricci tensor yields
\begin{equation}%7
\sqrt{-g} \; R_\mn (\G ) = \sqrt{-g} \; R_{\mn} (g) +
\textstyle{\frac{1}{2}}
\; g_\mn \; \pa_\a (h^{\a\b} X_\b ) \; .
\end{equation}
Since the extra term is a pure trace, it will not affect the
traceless $G_\mn$, which remains identically null.  Hence this
Palatini model is even more undetermined than its second order
form: not only is the metric left arbitrary, but so is $\G_\m$.
Note that the trace of (7),
\begin{equation}%8
h^\mn R_\mn (\G ) = \sqrt{-g} \; R(g) + \pa_\m (h^\mn X_\n ) \; ,
\end{equation}
shows that the scalar curvature density differs from the metric
Euler density by a divergence.  The second term on the right is
needed to cancel the Weyl dependence of the first, as is most
easily seen in a conformally flat frame, $g_\mn = e^{2\phi} \; \eta_\mn$,
in which the curvature depends only on
$\G_\m$:
\begin{equation}%9
h^\mn R_\mn (\G ) = - \pa^\m\G_\m
\end{equation}

We now turn to the more interesting  ``Kaluza--Klein reduced" model \cite{002}
involving a
Lagrange multiplier $N$.  The second order, metric, theory
$I = \int \; d^2x \; N \; h^\mn R_\mn (g)$ is no longer
vacuous, since the metric is now determined through the $R(g) =
0$ equation, while $N$ obeys
$D_\m  \pa_\n N = 0$.  One immediate difference is that unlike in metric form,
where a cosmological term is permitted in presence of $N$, the Palatini form
still excludes both $\l N \sqrt{g}$ and
$\l \sqrt{g}$ additions to ${\cal L}$: Weyl invariance again
forces $\l = 0$ here just as it did in the $N=1$ model.  In our
formulation, multiplying the
${\cal L}$ of (1) by $N$ now implies
\begin{equation}%10
h^\mn R_\mn (\G ) = 0
\end{equation}
as well as (2), and (3) with $h^\mn$ replaced by $N h^\mn$ there.
However, it is easy to see that $N$ must be constant: The trace
of the new (3) implies $D_\n (Nh^\mn ) =0$, so (4) holds with $N h^\mn$
replacing $h^\mn$; its $(\mn)$ trace reads
\begin{equation}%11
0 = g_\mn \, D_\a (N h^\mn ) \equiv g_\mn \pa_\a (
Nh^\mn ) \equiv
N g_\mn \pa_\a h^\mn + 2 \sqrt{-g} \, \pa_\a N \equiv 2\sqrt{-g} \: \pa_\a N \;
{}.
\end{equation}
Consequently, we may fall back on the previous results
of the $N=1$ model, except that now (10) is a field
equation, {\it i.e.}, (8) vanishes.  In conformal gauge, we see
from  (9) that $\G_\m$ is therefore divergenceless,
\be %12
\G_\m =  \e_\m~\!^\n \pa_\n \, \chi \; ,
\ee
but $\chi$ is still unrelated to the metric.  Furthermore, we have lost any
constraint on the metric,
since it is only the affine curvature scalar (8) that vanishes,
and that only contains $\G_\m$ as in (9), but not the
metric: This is again the legacy of Weyl invariance, that it
removes the one (conformal) variable in the metric tensor,
leaving nothing else to be determined.

Presence of matter does not alter things dramatically.  If it
does not involve the connection explicitly, the matter's stress tensor
defined according to $\d \, I_{MATT}/\d g_\mn$ will vanish since
$G_\mn$ does.  Although a (second quantized) spinor field action in 2D
is actually connection-independent, higher rank tensors will involve it
in general.  This dependence will introduce the usual matter torsion, but
not affect the metric indeterminacy of $\G_\m$.

Very similar considerations hold when zweibeins $e_{\m a}$/connections $\o_{\m
ab}$ are
used instead of the metric.  The Einstein Lagrangian here
($e \equiv | \det e_{\m a}|$ and $e^{\m a}$ is the usual inverse)
\begin{eqnarray}%13
{\cal L} & = &   e  e^{\m\a}e^{\n b} R_{\mn ab}(\o ) \; ,
\;\;\;
\o_{\m ab} \; =  - \o_{\m ba} \nonumber \\
R_{\mn ab } (\o ) & \equiv & (\pa_\m \o_{\n ab} - \o_{\m ac}
\o_{\n cb} ) - (\n\m )
\end{eqnarray}
is still Weyl-invariant at D=2, since it is homogeneous of order zero in the
zweibeins, and also
simplifies drastically, since we may write $\o_{\m
ab} \equiv
\e_{ab} \, \o_\m$, thereby reducing $R_{\mn ab} $ to the abelian
form
$\e_{ab}(\pa_\m \o_\n - \pa_\n \o_\m )$ and ${\cal L}$ to the minimal
expression
\begin{equation}%14
{\cal L} = 2\e^\mn \pa_\m \o_\n
\; ,
\end{equation}
in terms of the (constant) Levi--Civita  density $\e^\mn$.  Thus,
the first order theory does not involve the
zweibein $e_{\m a}$ at all, let alone determine $\o_\m$ in terms of it.
Indeed, there are no field equations at all here!  The ``K-K reduced" theory,
multiplying $\cal L$ by $N$
only implies that
$\e^\mn \pa_\m \o_\n = 0$ and $N =$ const.  Again, Weyl invariance requires
that  $\l =0$  for either $\l e$ or $\l Ne $ cosmological terms.

Our final ``different in D=2" model is an old formulation of Einstein
gravity, due to Eddington \cite{003}.  His
proposal was to consider the purely affine Lagrangian
\begin{equation}%15
{\cal L}_E = (- \det \, R_\mn (\G ))^{1/2} \; ,
\end{equation}
in terms of the symmetrized
part of the
$R_\mn (\G )$ in (1).  Since $R_\mn$ is a tensor, ${\cal L}$ is a scalar
density and the field equations are tensorial, resembling (4):
\begin{equation}%16
D_\a ( R^\mn \sqrt{-\det R_\mn} ) = 0
\end{equation}
where $R^\mn$ is the (assumed to exist)  matrix inverse of $R_\mn$.  This
follows
simply from the fact that for any determinant, $\d (\det \, R_\mn
) = R^\mn \d R_\mn (\det \, R_\mn )$, and from the
(symmetrized) Palatini identity
$\d R_\mn = D_\a \, \d \, \G^\a_\mn - \textstyle{\frac{1}{2}}
(D_\m \, \d \G_\n + D_\n \d \G_\m )$.  In general dimension, (16)
means that
$R^\mn (\G ) \sqrt{-\det R_\mn}$, and hence also $R_\mn (\G )$, is
covariantly constant, {\it i.e.}, if we call $R_\mn (\G )$ by the name
$g_\mn$,
$$%17a
R_\mn (\G ) = \l g_\mn \; , \eqno{(17a)}
$$
then $g_\mn$ is covariantly constant
$$%17b
D_\a (\G ) g_\mn = 0 \; . \eqno{(17b)}
$$
But these are of course the Einstein equations for the metric $g_\mn$ with a
cosmological  term.  This reasoning fails
precisely at D=2 because $R^\mn   \sqrt{-\det R_\mn} $ is
unimodular (for any 2D symmetric tensor!) and so (as we have seen in detail
earlier) there are not enough variables available in (16) to
specify the \underline{full} metric, {\it i.e.}, to imply (17).
We are again reminded that a seemingly generic statement like (16) can
degenerate in a particular dimension.

We have not investigated whether the ``ultratopological" models discussed here
might have interesting quantum consequences.

This work was supported by the NSF under grant \#PHY--9315811.

\end{document}